# Review of Magnetic Shark Deterrents: Hypothetical Mechanisms and Evidence for Selectivity

Joshua Courtney, Ya'el Courtney, Michael Courtney
BTG Research
Baton Rouge, Louisiana, USA

**Abstract**
Several papers published since 2006 describe effects of magnetic fields on elasmobranchs and assess their utility in reducing negative interactions between sharks and humans, including bycatch reduction. Most of these repeat a single untested hypothesis regarding physical mechanisms by which elasmobranchs detect magnetic fields and also neglect careful consideration of magnetoreception in teleosts. Several species of teleosts are known to have magnetoreception based in biogenic magnetite, and direct magnetic field detection also has support in several species of elasmobranchs. The overly narrow focus of earlier papers on the unsupported hypothesis that magnetoreception in elasmobranchs is based in the ampullae of Lorenzini creates the impression that all teleosts will be insensitive to magnetic deterrents. However, magnetite based magnetoreception has been demonstrated in several teleosts, and is supported in others. Furthermore, electroreception is present in many teleost species; therefore, the possibility of induction based indirect magnetoreception should be considered. Finally, experiments reported as demonstrating insensitivity in teleost species to magnetic deterrents suffer from inadequate design and sample sizes to reject the hypothesis of magnetic detection in any given species. Since adoption of deterrent hook technologies depends on both deterrent effects in sharks and the absence of effects in target teleosts, the hypothesis of detection in teleost species must be independently tested with adequate sample sizes.

## 1. Introduction and Background

Magnetoreception is the capability to detect a magnetic field and, in some cases, use it to guide movement. Over the last century, it has been hypothesized and then supported experimentally that many species have a magnetic sense. Evidence has been found to support magnetoreception in certain species of bacteria, mollusks, insects, amphibians, birds, reptiles, mammals, and fish. (Johnsen & Lohmann, 2004) In principle, the Earth's magnetic field provides positional information that may be used for orientation and navigation. There are geometric parameters such as field intensity and inclination of field lines relative to Earth's surface that vary with location. Evidence has accumulated steadily that animals can derive positional information from these parameters. (Kirschvink, 1989; Walker et al., 2003; Walker et al., 2006; Hart et al., 2012)

One group of animals that has been shown to exercise magnetoreception is elasmobranch fish, including sharks and rays. Sharks are also part of a large group known as bycatch – marine animals that are caught while fishing for another species. Shark bycatch contributes to population declines and management difficulties, as well as inefficiency in commercial fisheries. (Godin et al., 2013) Because high shark bycatch can result in bait loss, gear damage, and risks to fishing crew, (Gilman et al., 2008) reducing shark interactions is a priority for fisheries managers. There are various approaches to reducing shark interactions, but this review will focus on magnetic deterrents and hypothetical sensory mechanisms, evidence for each possible mechanism, and species selectivity relating to potential use of magnetic deterrents.

Several papers describe effects of magnetic fields on elasmobranchs and purport to assess their utility in reducing negative interactions between sharks and humans. (Godin et al., 2013; O'Connell et al., 2011; O'Connell et al., 2013; Tallack & Mandelman, 2009; Jordan et al., 2013; Stoner & Kaimmer, 2008; Robbins et al., 2011) However, most of these papers discuss only a single, untested hypothesis regarding magnetic field detection mechanisms in elasmobranchs, consequently neglecting adequate consideration of magnetoreception in teleost fish. This hypothesis focuses on electromagnetic induction with the ampullae of Lorenzini as the detection organ and detecting induced electric fields as the mechanism. This creates the expectation that all teleosts are insensitive to magnetic deterrents. Hypotheses should be tested explicitly without relying on unverified theoretical extrapolations. Numerous experiments have given support to electroreception in teleosts. (Northcutt et al., 1995; Modrell et al., 2011a ; New, 1997; Bodznick & Northcutt, 1981; Jorgensen, 2005) Several studies have also shown evidence of direct magnetoreception through biogenic magnetite in teleosts and several elasmobranch species, suggesting that due consideration should be given to alternate mechanistic hypotheses. (Kirschvink et al., 2001; Johnsen & Lohmann, 2004; Walker et al., 2006; Eder et al., 2012) Thus, further experiments are needed regarding magnetoreception through biogenic magnetite and magnetoreception in teleosts.

## 2. Transduction Hypotheses

There are several hypotheses regarding mechanisms of magnetic reception in fish. One with strong support is direct magnetic field detection, (Kirschvink et al., 2001; Johnsen & Lohmann, 2004) which asserts that direct detection is based on the magnetic mineral, magnetite ($Fe_3O_4$). (Walker et al., 2006) It proposes that fish use motion or torque from magnetite crystals to convert magnetic field stimuli into mechanical signals that are detected by the nervous system. (Kirschvink, 1989; Eder et al., 2012) These crystals are permanently magnetized bar magnets that twist into alignment with an externally applied magnetic field if allowed to rotate freely. In many fish, these are under 50 nm, with the exception of the spiny dogfish's otolith, which is larger.

A second hypothesis regarding magnetic reception is electromagnetic induction. (Johnsen & Lohmann, 2004; Albert & Crampton, 2006) In general, a conductor moving in a magnetic field or a magnet moving through a conducting medium (salt water) produces an induced electric field. The induced electric field is proportional to the strength of the magnetic field and to the relative velocity between the conductor and the magnetic field. (Walker et al., 2003; Lohmann et al., 2008) This hypothesis deals primarily with fish and requires fish with this sense to have an electroreceptive organ system that detects an externally applied electric field. This electric field may be created by a stationary magnetic field in a moving saltwater current. However, some authors propose that the organism itself (and its sensory systems) complete the circuit. This view favors the ampullae of Lorenzini as the sensory organ in elasmobranchs and various electroreceptive organs in bony fish. (Johnsen & Lohmann, 2004; Gillis et al., 2012)

Chemical magnetoreception is a third hypothesis regarding magnetic reception but will not be discussed further in this review because there is insufficient evidence that this is relevant in fish. This hypothesis requires chemical reactions that are affected by magnetic fields comparable in magnitude to the Earth's magnetic field (~ 50 gauss). No biological reactions have been identified that completely fulfill the properties required. (Johnsen & Lohmann, 2004)

## 3. Evidence for Magnetoreception Mechanisms

### 3.1 Direct Magnetic Reception

Biogenetic magnetite has been shown to be present in several species of teleosts that demonstrate magnetoreception including rainbow trout (*Oncorhynchus mykiss*), yellowfin tuna (*Thunnus albacares*), chinook salmon (*Oncorhynchus tshawytscha*), and sockeye salmon (*Oncorhynchus nerka*). (Driebel et al., 2000; Walker et al., 1984; Kirschvink et al., 1985; Mann et al., 1988) There is also compelling evidence for the presence of direct magnetoreception in swordfish (*Xiphias gladius*), based on its ability to navigate long distances along a given compass heading without any other plausible explanations for maintaining a given orientation. (Carey & Robinson, 1981) Behavioral evidence for magnetoreception is present in the Japanese eel (*Anguilla japonica*), American eel (*Anguilla rostrata*), Atlantic salmon (*Salmo salar*), zebrafish (*Danio rerio*), Mozambique tilapia (*Oreochromis mossambicus*), common carp (*Cyprinus carpio*) and other species. (Nishi et al., 2004; Rommel & McCleave, 1973; Shcherbakov et al., 2005; Hart et al., 2012)

With regard to direct magnetoreception in elasmobranch fish, there is evidence for direct magnetoreception in scalloped hammerhead shark (*Sphyrna lewini*) and short-tailed stingray (*Dasyatis brevicaudata*), which casts doubt on implicit suggestions that magnetic reception is always mediated by electromagnetic induction through the ampullae of Lorenzini. (Klimley, 1993; Hodson, 2001) Magnetite has also been found in the stataconia of the spiny dogfish (*Squalus acanthias*). (Hanson et al., 1990) Several authors have argued that electroreception as a mechanism for detecting the earth's magnetic field would be inefficient because all elasmobranchs in question lack structures of appropriately large size to achieve necessary magnetic sensitivity. (Rosenblum et al., 1985; Semm & Beason, 1990; Klimley, 1993) Observed behavior in sharks and rays also indicates that the primary purpose of electroreception is to locate prey. (Hodson, 2001; Walker et al., 2003; Kirschvink et al., 2001) Based on the above findings, Kirschvink, Walker, and Diebel (2001) concluded that experimental evidence rules out electroreception as the basis of magnetoreception in elasmobranchs.

Several studies regarding magnetic reception in elasmobranch fish ascribe effects to indirect magnetoreception without experimental evidence to rule out magnetite based magnetoreception.

It has been shown that many elasmobranch species have capability to detect magnetic fields. This includes juvenile nurse sharks and lemon sharks, as well as many species of mature sharks. (O'Connell et al., 2011; O'Connell et al., 2010) Further experimentation is needed in confirmed cases of magnetoreception to determine the mechanism.

*3.2 Magnetoreception by Electromagnetic Induction*

A conductor moving in a magnetic field generates electric fields proportional to the magnetic field strength and speed of motion. Thus, any electrosensitive organism can potentially detect a magnetic field if the combination of speed and magnetic field strength creates a detectable electric field. Speed can be generated by moving salt water (current), a moving magnet, or a moving fish.

One hypothesis regarding electroreception in bony fish points to ampullary organs as the mechanism. It is a long established observation that numerous species of bony fish possess electroreceptive capabilities. (Kramer, 1996; Bretschneider & Peters, 1992; Albert & Crampton, 2006) Electroreceptors are housed in ampullary sense organs. (Jorgensen, 2005) They include epidermal hair cell receptors, receptor organs in which hair cells extend into a fluid-filled lumen. Ampullary receptor organs in non-elasmobranch fish differ in several ways from those of marine elasmobranchs: the canals are shorter, there are fewer hair cells per organ, and there is usually only a single afferent fiber from each organ. (Albert & Crampton, 2006) Non-teleost fish that possess both ampullary organs and electroreceptive capabilities include paddlefish (*Polyodon spathula*), Australian lungfish (*Neoceratodus forsteri*), coelacanths (*Latimeria chalumnae*), and various sturgeon species. Teleost fish that possess both ampullary organs and electroreceptive capabilities include many species of catfish, African mormyriformes, and neotropical gymnotiformes. (Bullock et al., 1983; Hopkins, 1995)

Bony fish may also sense electric fields by other mechanisms. Some fish use tuberous electroreceptors to detect electric fields, including those generated by electrogenic fish. (Albert & Crampton, 2006) Tuberous organs are similar to ampullary organs, with electrosensory hair cells and a canal extending to a superficial pore. The core difference between tuberous and ampullary organs is that in tuberous organs, the hair cells lie mostly within the organ lumen. The electrosensory system of the lamprey consists of small swellings called end buds distributed on the epidermal surface over the whole body. (Ronan, 1986)

Authors on elasmobranch electroreception hypothesize that the ampullae of Lorenzini serve as the location of detection. (O'Connell et al., 2012; Albert & Crampton, 2006; Wueringer et al., 2012; Johnson & Lohmann, 2004) In elasmobranchs, ampullary organs are clustered into discrete regions on the head and pectoral fins, and their canals connect with pores distributed on the surface of the skin. It is hypothesized that this organization allows ampullary electroreceptors to detect potential differences between a common internal potential and the somatotopic charges on the skin. (Bleckmann, 1994)

The findings of Bullock and Northcutt (1982) reveal the possibility that "electroreception might turn up anywhere among hundreds of fish families, especially among teleosts … it will not necessarily be homologous to previous known examples." (Bullock, 1999) Electroreception and insensitivity to the presence of permanent magnets should not be inferred for broad classes of teleosts based on theoretical considerations or limited data since "most of the 30 orders of fishes not known to have electroreception have probably not been adequately examined … the task is much larger than sampling 30 orders." (Bullock & Northcutt, 1982) This suggests the possibility of indirect magnetoreception in almost any species of fish and that insensitivity to electric fields should be tested explicitly and not assumed based on unverified theoretical inferences.

## 4. Magnetic Deterrents

Early efforts to develop shark deterrents were motivated by the need to protect humans in shark infested waters. More recent efforts have been motivated by conservation needs to reduce elasmobranch mortality associated with bycatch of fisheries and beach nets. (Bonfil, 1994; Shepherd & Myers, 2005) One technology commonly evaluated as a shark repellent is permanent magnets. The use of magnetic hooks has only a nominal cost increase, making magnetic hooks commercially viable if they are effective at the deterrent function without reducing catch rates of target species. These magnets are hypothesized to work by overstimulating the ampullae of Lorenzini that are present in elasmobranchs. (Stoner & Kaimmer, 2008; Jordan et al., 2013; O'Connell et al., 2012; Rigg et al., 2009) Because teleosts have not been shown to possess ampullae of Lorenzini, proponents of this hypothesis have attempted to infer that magnetic repellents selectively repel elasmobranchs but not teleosts. Since viable deterrent technologies must be selective to reduce bycatch of elasmobranchs in fisheries without reducing catch of target species, it is essential to consider the strength of the theoretical and empirical evidence regarding whether or not teleost species are likely to be sensitive to deterrent technologies.

Permanent magnets on hooks have been experimentally shown to reduce the catch of some elasmobranch species. Hook-and-line experiments have supported that magnets reduce the catch rates of the Atlantic sharpnose shark (*Rhizoprionodon terraenovae)* and the smooth dogfish (*M. canis*). (O'Connell et al., 2011) Longline experiments found that permanent magnets significantly decreased capture of the blacktip shark (*Carcharhinus limbatus*) and the southern stingray (*D. Americana*). (O'Connell et al., 2011)

Very few statistically significant results have been published regarding whether magnetic hooks alter catch rates in teleost species. Because of the current reliance on theoretical considerations, adequate tests of the hypothesis that permanent magnets do not alter teleost capture are required. Some current research on this topic has presented inadequate study designs to reach significant conclusions. A 2011 paper featured a secondary hypothesis regarding teleost capture on hooks with permanent magnets, but used a sample of only four teleosts in a longline group and eleven in a hook-and-line group to assert conclusions. (O'Connell et al., 2011; Courtney & Courtney, 2011) Whether or not magnetic hooks reduce catch rates of teleost fishes remains largely an open question. It is necessary to explicitly test this hypothesis with adequate sample sizes for a number of taxa of teleost fish under varying conditions.

## 5. Discussion

There are two main hypotheses regarding the mechanism of magnetoreception in fish: direct and indirect. Direct magnetoreception involves biogenic magnetite, and there is evidence for this in both elasmobranch and teleost fish. Indirect magnetoreception involves electromagnetic induction. It is well established that many fish are electroreceptive and have specific reception mechanisms that may be used to detect magnetic fields as well. Elasmobranch fish detect electric fields through ampullae of Lorenzini, while other bony fish detect electric fields through ampullary organs or other mechanisms. Direct electroreceptive capabilities in a fish do not exclude the possibility of direct magnetoreceptive capabilities also. Further, it should not be assumed that fish without known electroreceptive capabilities are insensitive to electric and magnetic fields without testing explicitly.

In any given species, detecting significantly different catch rates between a magnetic hook and a sham would suggest some form of magnetoreception in that species. Sensitivity to magnetic hooks may be the simplest method yet proposed for detecting magnetic sensitivity in fish, and experiments can be conducted relatively inexpensively in field trials in all species available to be readily caught on baited hooks. This is much simpler and more accessible than previous methods involving various conditioning techniques and complex laboratory apparatus. Further, field trials with magnetic hooks allow study of magnetic sensitivity in species prohibitively large, difficult, or expensive to manage under captive laboratory conditions.

If magnetic reception exists only indirectly via electromagnetic induction, sensitivity to magnetic hooks will likely vary between still and moving water. Since the induced electric field is proportional to the current speed, fish that detect magnetic fields indirectly will lose sensitivity in still water, although the movements of the fish may provide a minimum amount of motion needed to detect magnets at sufficiently close distances. In contrast, fish that detect magnetic fields by biogenic magnetite should be sensitive to magnetic hooks without regard for the presence of current.

Since aquatic conductivity is necessary to complete the circuit in fish that detect magnetic field via electromagnetic induction, it is likely that species whose magnetic sensitivity depends strongly on salinity levels (thus aquatic conductance) are making use of electromagnetic induction. Insensitivity of magnetic detection thresholds to salinity over a broad range would suggest direct magnetoreception is more likely.

Sensitive techniques have been developed to identify biogenic magnetite in magnetoreceptive fishes, and some authors (Kirschvink et al., 2001) have gone as far as to assert that the presence of magnetite is the defining feature of magnetoreceptive vertebrates. A confirmed absence (using the most sensitive available techniques) of biogenic magnetite would suggest that indirect electroreception is the mechanism at work in magnetoreceptive species. Conversely, confirmation of the presence of biogenic magnetite in any given species suggests that direct magnetic field detection is present in that species.

The necessary selectivity of magnetic shark deterrents has been assumed though flawed theoretical reasoning that teleost species cannot detect electric and magnetic fields. Given that both electroreception and direct magnetoreception have been demonstrated in broad classes of teleosts, attempting to establish teleost insensitivity to magnetic deterrents from results in small numbers of teleost taxa or by grouping of all teleosts is unwarranted. The hypothesis of teleost insensitivity to magnetic hooks should be tested with adequate sample sizes for a number of taxa under varying conditions including moving water. Failure to explicitly test for sensitivity to magnetic deterrents creates a real risk of inadvertently altering catch rates of teleost species.

## 6. Acknowledgments

The subject of magnetoreception in fish first came to our attention when we read O'Connell et al. (2011) and noted the numerous errors. We brought these to the authors' attention, and while the data and analysis reporting errors were corrected in a published *errata*, the scientific errors remained and continued to be propagated in additional publications. The journals propagating these errors refused to balance the information with replies citing Kirschvink and his co-authors regarding magnetite based magnetoreception, with one editor telling us, "I could not find that you ever contributed to this sort of research." We contacted Joe Kirschvink at Cal Tech for advice. He described the editor's obvious bias as "idiotic" and encouraged us to persist, which has resulted in this paper. We are grateful to BTG Research and the United States Air Force Academy for funding this review paper as well as original research in which we have documented magnetoreception in three species of teleosts: black drum (*Pogonias cromis*), sea catfish (*Ariopsis felis*), and red drum (*Sciaenops ocelatus*). Though the corresponding author (MC) is no longer affiliated with the Air Force Academy, he is grateful to their support and encouragement allowing him to explore and enter new research fields. We are also grateful to the two anonymous peer reviewers (Aquatic Science and Technology) who provided valuable feedback which was incorporated into the paper.